\documentclass[adp,fleqn]{w-art}
\usepackage{times}
\usepackage{w-thm}

\usepackage{amssymb,amsmath}% mathpackage
\usepackage{ushort}% for the underlines and double underlines of operators
\usepackage{bbm}
\usepackage{rotating}% sidewaysfigures
\usepackage{pgf}% sidewaysfigures

\renewcommand{\cal}[1]{\ensuremath{\mathcal{#1}}}

\usepackage[]{graphicx}
\begin{document}
%%    The information for the title page will be placed between
%%    \begin{document} and \maketitle. The order of most entries
%%    is determined by the class file and can not be changed by
%%    rearranging them. The maketitle command follows after the
%%    absract.
%%
%%    Most of the following commands will be completed by the publisher.
%%
%%    The copyrightyear is defined in the .clo file as the first argument
%%    of the copyrightinfo command. If the copyrightyear differs from that
%%    value it might be adjusted by the following definition:
%%
%% \renewcommand{\copyrightyear}{2002}% uncomment to change the copyrightyear.
%%
%\DOIsuffix{theDOIsuffix}
%%
%% issueinfo for header and copyright line
%\Volume{16}
%\Issue{1}
%\Copyrightissue{01}
%\Month{01}
%\Year{2007}
%%
%%    First and last pagenumber of the article. If the option
%%    'autolastpage' is set (default) the second argument may be left empty.
\pagespan{3}{}
%%
%%    Dates will be filled in by the publisher. The 'reviseddate' and
%%    'dateposted' (Published online) entry may be left empty.
%\Receiveddate{15 November 2007}
%\Reviseddate{30 November 2007}
%\Accepteddate{2 December 2007}
%\Dateposted{3 December 2007}
%%
\keywords{dynamical mean field theory, {\em ab initio} calculations, many body theory}

%% \pretitle{Editor's Choice}

%% We have a short and a long form for the title. The short form
%% (optional argument) goes into the running head.

\title[Ab initio D$\Gamma$A]{Ab initio calculations with the dynamical vertex approximation}

%% Please do not enter footnotes or \inst{}-notes into the optional
%% argument of the author command. The optional argument will go into
%% the header.  If there is only one address the marker \inst{x} may be
%% omitted.

%% Information for the first author.
\author[]{A. Toschi\inst{1}}%
\address[\inst{1}]{Institute of Solid State Physics, Vienna University of Technology, 1040 Vienna, Austria}
%%
%%    Information for the second author
\author[]{G. Rohringer\inst{1}}
%%
%%    Information for the third author
\author[]{A. A. Katanin\inst{2}}
\address[\inst{2}]{Institute of Metal Physics, 620219 Ekaterinburg, Russia}

%%
%%    Information for the third author
\author[]{K. Held\inst{1}
  \footnote{Corresponding author\quad E-mail:~\textsf{held@ifp.tuwien.ac.at}, 
            Phone: +43\,1\,58801\,13710}}

\begin{abstract}
We propose an approach for the {\em ab initio} calculation
of materials with strong
electronic correlations which is based on all local (fully irreducible)
vertex corrections beyond the bare Coulomb interaction. It includes the so-called $GW$ and dynamical mean field theory and
important non-local correlations beyond, with a computational effort estimated
to be still manageable.
\end{abstract}
%% maketitle must follow the abstract.
\maketitle                   % Produces the title.

%% If there is not enough space inside the running head
%% for all authors including the title you may provide
%% the leftmark in one of the following three forms:

%% \renewcommand{\leftmark}
%% {F. Author: A short title}

%% \renewcommand{\leftmark}
%% {F. Author and S. Author: A short title}

%% \renewcommand{\leftmark}
%% {F. Author et al.: A short title}

%% \tableofcontents  % Produces the table of contents.

\section{Introduction}
Twenty-two years ago Metzner and Vollhardt \cite{Metzner89a} 
initiated dynamical mean field theory (DMFT) by investigating
fermionic lattice models in the limit of infinite 
dimensions ($d=\infty$). In this limit, the (irreducible)  Feynman diagrams for
the self energy greatly simplify to their local contribution 
 \cite{Metzner89a},
and non-local interactions 
reduce to their Hartree contribution
\cite{MuellerHartmann89a}. Subsequently,
 Georges and Kotliar \cite{Georges92a}  showed that fermionic lattice models are, for $d=\infty$,
mapped    onto the self-consistent solution of an auxiliary 
Anderson impurity model, a most crucial step since this allowed
for using well-known impurity solvers. As a direct consequence, the Mott
transition \cite{Georges92a}  and antiferromagnetism  \cite{Jarrell92a}
were  studied  by DMFT.

In the last years, 
DMFT \cite{Georges96a,Kotliar04a} along with its cluster \cite{DCA,clusterDMFT,LichtensteinDCA,Maier04}  and diagrammatic \cite{Schiller,Toschi06a,Kusunose06a,Slezak06a,DualFermion} extensions
became one of the standard approaches for calculating strongly correlated electron
systems. It is employed
 not only for model Hamiltonians but combined with density functional theory in
the local density approximation (LDA) \cite{Jones89a} also for the
realistic calculation of  material properties \cite{Anisimov97a,Lichtenstein98a,Held03,Kotliar06,Held07a}. Despite its success, merging LDA with DMFT has severe drawbacks as density functional theory and many body Feynman diagrams are rather orthogonal approaches which do not match nicely. Most noticeable is the so-called double counting problem, arising from the fact that it is unclear how to express, in the DMFT  language of Feynman diagrams, the 
correlations already included in the exchange correlation potential of LDA. But also the calculation of the local Coulomb interaction 
needed for the DMFT calculation is problematic since DMFT does not include non-local screening effects. These must be taken into account already through a reduced $U$ value,
e.g., calculated by constrained LDA  \cite{Dederichs84,McMahan88,Gunnarsson89}
or by the constrained random phase approximation (cRPA) \cite{cRPA1,cRPA2}

\begin{figure}
\begin{center}
\includegraphics[clip=true,width=11.5cm]{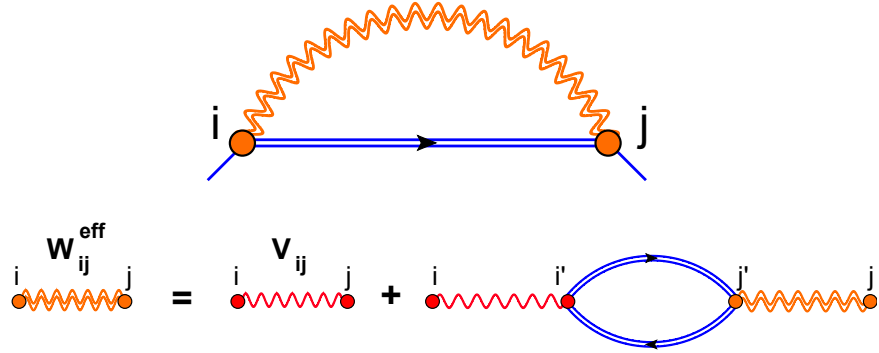}
\end{center}
\vspace{.6cm}

\caption{{\em Top:} 
In addition to the  Hartree term, the $GW$ approach takes into
account the (screened) exchange diagram (single wiggled line: bare Coulomb interaction; double wiggled line: screened Coulomb interaction; double line: interacting Green function; here and in the following $i,j$ indicate the lattice sites and also subsume the spin indices).\newline
{\em Bottom:} Screening of the Coulomb interaction within
the random phase approximation (RPA) .\label{Fig:GW}}
\end{figure}

\begin{figure}
\begin{center}
\includegraphics[clip=true,width=14cm]{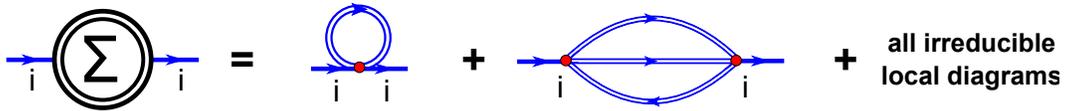}
\end{center}
\caption{Feynman diagrams for the DMFT self energy $\Sigma$ in terms of the interacting Green function $G$ (double line). (Here and in the following, we omit the single wiggled line, i.e., use only a dot,
if only the bare local interaction is taken. Single wiggled lines are kept if also non-local interactions are included.)
\label{Fig:DMFT}}
\end{figure}

A way to overcome these two obstacles is to substitute LDA by Hedin's so-called $GW$ approach \cite{Hedin}, which generally gives similar results as LDA but with an improved description of the exchange. Hence semiconductor band gaps are much better described, see, e.g., \cite{Hybertsen85} and for more recent
results \cite{P7:ShishkinPRL07}.
This $GW$+DMFT approach \cite{P7:GW+DMFT} is conceptionally clear  and
well-defined in terms of Feynman diagrams, see Figures \ref{Fig:GW}, \ref{Fig:DMFT}.
It accounts for the screening of the Coulomb interaction and being determined
entirely in terms of Feynman diagrams the subtraction of double
counted diagrams becomes a well defined problem.
Despite these advantages,  the application and further development of
the $GW$+DMFT approach  has been stalled since its formulation by Biermann {\em et al.} \cite{P7:GW+DMFT}. A first reason for this is that
the $GW$ approach
is computationally fairly demanding and complex. Hence,
mature $GW$ programs were missing in the past, albeit this changed more recently, see e.g.\ \cite{P7:Faleev04,P7:Chantis06,P7:Shishkin06,P7:Scheffler08}. Second,
the $GW$+DMFT scheme is considerably more involved than LDA+DMFT, in particular,
if calculations are done self-consistently and with a frequency dependent
(screened) Coulomb interaction. In this situation a full implementation of $GW$+DMFT appears to be  feasible, albeit with quite a considerable effort.

\begin{figure}
\begin{center}
\includegraphics[clip=true,width=13cm]{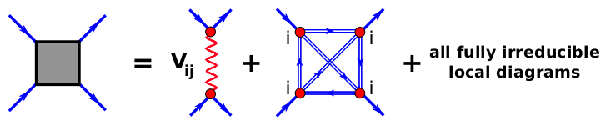}
\end{center}
\vspace{.6cm}
\caption{{\em Ab initio dynamical vertex approximation:} \newline
Besides the bare  Coulomb interaction at the lowest order, all local vertex corrections to  the fully irreducible vertex $\Gamma_{\rm ir}$ 
(filled square box) are
included. This fully irreducible vertex is calculated numerically from an Anderson impurity model, and is used as input for
the parquet equations (see Fig. \ref{Fig:parquet}).  \label{Fig:DGA}}
\end{figure}

On the other hand, from a mere theoretical perspective,
 $GW$+DMFT is a rather ad-hoc merger of two (in their respective field successful)  approaches  without an underlying principle. In this paper, we point out that {\em ab initio} calculations with dynamical vertex approximation (D$\Gamma$A) is a common basis which includes
the $GW$ and DMFT diagrammatic contributions as well as non-local correlations 
beyond in a natural way. {\em Ab initio} D$\Gamma$A also realizes
Hedin's original ideas \cite{Hedin} to include vertex corrections beyond $GW$. Not all vertex corrections are included, which is of course not possible. Instead,
only those corrections are incorporated 
which can be traced back to a fully irreducible local 
vertex, see Fig.\ \ref{Fig:DGA}. This way we obtain a purely Feynman diagrammatic approach 
which, while being  obviously more involved than $GW$+DMFT, is computationally 
still feasible. It should  describe weakly correlated systems similarly
good as LDA and even better if exchange is important as in semiconductors. It includes the strong local DMFT correlations which are so important for quasiparticle renormalizations, Mott-Hubbard transitions, magnetic moments etc. Finally, it also includes non-local correlations beyond DMFT which provide for the most fascinating aspects of correlated electrons such as 
spin fluctuations and their interplay with superconductivity, quantum criticality and
weak localization.

In the following, we will discuss how to do {\em ab initio} calculations with the D$\Gamma$A for actual materials.
In Section \ref{Sec:DGA} we introduce the approach in terms of Feynman diagrams which can be translated one-to-one to equations. Let us emphasize, our paper is not a review on previous
 D$\Gamma$A calculation \cite{Toschi06a,Kusunose06a,Slezak06a,Katanin09a,Rohringer11a}
but instead points out how to do calculations for actual materials.
Moreover,  in Section  \ref{Sec:GW}  and  \ref{Sec:DMFT},
we show explicitly  that $GW$ and DMFT, respectively,
are included,   as are non-local correlations beyond. 
 Section \ref{Sec:algorithm} summarizes the algorithm of the proposed scheme and
discusses the effort of the algorithm. Finally,
Section  \ref{Sec:outlook} gives a summary and outlook.

\section{Ab initio dynamical vertex approximation (D$\Gamma$A)}
\label{Sec:DGA}
The basic idea of D$\Gamma$A is to put the DMFT concept of taking 
all local contributions of the Feynman diagrams for the self energy (one particle irreducible vertex) to a higher level. That is, requiring the fully 
irreducible $n$-particle vertex to be local. This approach has been put
forward by some of us \cite{Toschi06a} and, taking a cluster of sites
as a starting point, by Slezak {\em et al.} \cite{Slezak06a}; also note related ideas by Kusonose
\cite{Kusunose06a}. While $n\rightarrow \infty$ yields the exact solution
of a given fermionic Hamiltonian, most promising is the $n=2$ level as 
this only requires the calculation of two particle response functions
(not $n$ particle ones) for an Anderson impurity model
 \cite{Toschi06a,Held08a}, a very feasible task. Nonetheless,
on this $n=2$ level most (if not all) of the known correlation physics
is already included.
The approach was shown to correctly describe pseudogaps \cite{Katanin09a},
the Mermin-Wagner theorem in two dimensions \cite{Katanin09a}, and critical exponents in three dimensions \cite{Rohringer11a}. Its implementation for nanoscopic systems
is intriguing as it includes important vertex and weak localization corrections which are important for the conductance \cite{Valli10a}.
D$\Gamma$A is closely related to the dual Fermion approach \cite{DualFermion} which considers similar diagrammatic expansions, albeit not for the actual Fermions but for dual Fermions. In this sense, the connection between 
D$\Gamma$A and dual Fermions is comparable to that of dynamical cluster approximation \cite{DCA} and cluster DMFT \cite{clusterDMFT} which consider similar clusters in $k$-space and real space, respectively.

As already mentioned, in {\em ab initio} D$\Gamma$A 
the fully irreducible vertex is approximated by its
local contribution ($k$-independent but fully frequency dependent) 
except for for the lowest order term, i.e., the bare interaction, which is taken with its full spatial dependence see Figure \ref{Fig:DGA}:
\begin{equation}
 \Gamma_{ij \; \rm ir}  =  V_{ij}  + \mbox{all local vertex corrections}
\label{Eq:Gamma}
\end{equation}

\begin{figure}
\begin{center}
\includegraphics[clip=true,width=14cm]{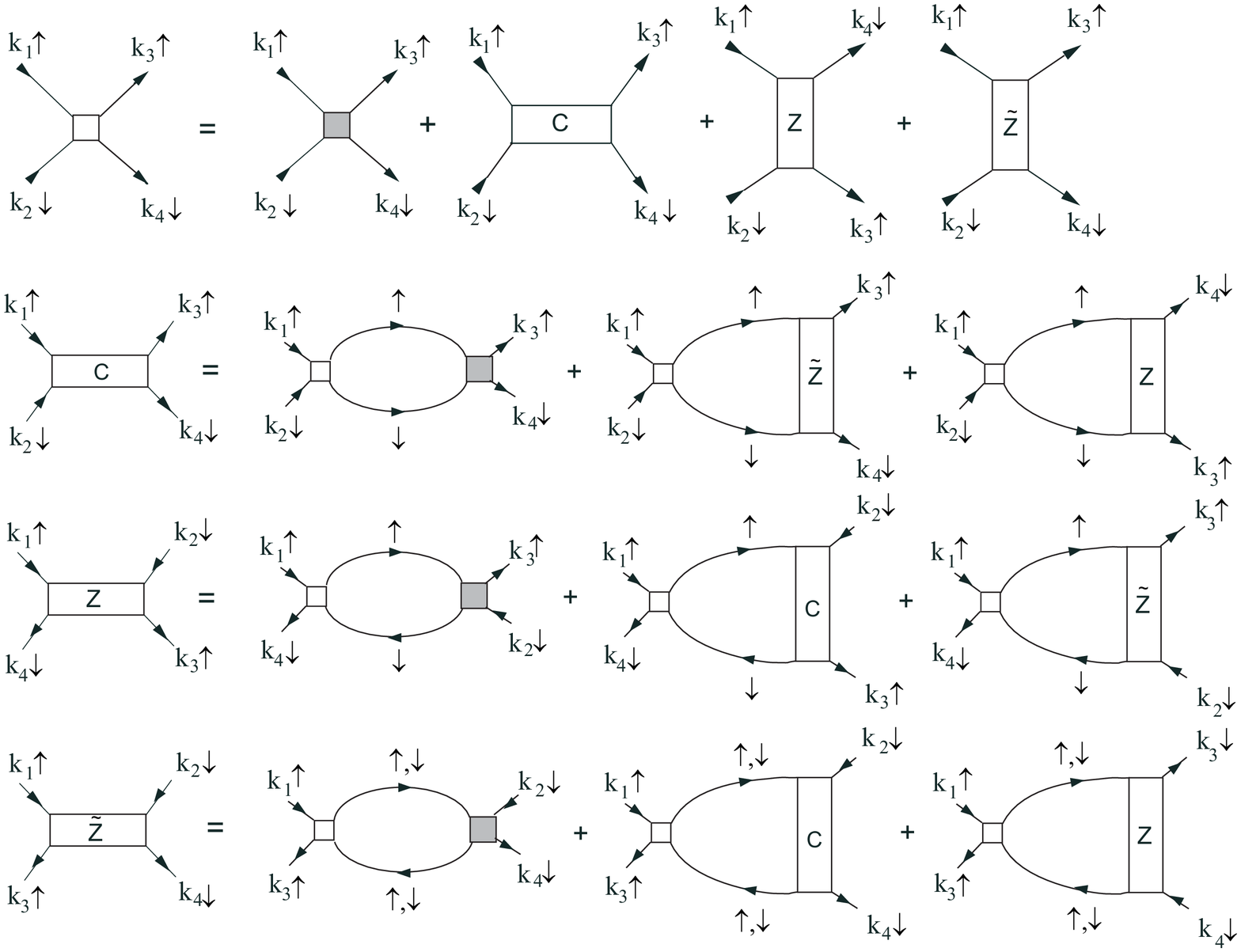}
\end{center}

\caption{{\em Parquet equations} connecting the reducible vertex $\Gamma$ (empty squares), the fully irreducible vertex $\Gamma_{\rm ir}$ (filled squares) and the reducible vertex
in the particle-particle  ($C$) and two particle-hole channels ($Z$, $\tilde Z$)
\label{Fig:parquet} (see \cite{Toschi06a} for the corresponding equations).}
\end{figure}

From this fully irreducible vertex $\Gamma_{\rm ir}$ in turn, the reducible vertex $\Gamma$ is calculated via the
parquet equations \cite{Dzy,parquet,Janis2}, see Figure \ref{Fig:parquet}, or, neglecting certain channels (e.g., the particle-particle channel), with the corresponding the Bethe-Salpeter equations. 
While we have been concentrating in the past more on the latter, i.e., the  approximative Bethe-Salpeter calculation of the reducible vertex, including a Moriya-like $\lambda$-correction \cite{Katanin09a},  Jarrell and coworkers made remarkable progress in
actually solving the parquet equations on a $k$-grid \cite{Yang09a}.
Hence, a parquet solution is nowadays possible not only for a frequency independent interaction (parquet approximation) but also for a fully frequency dependent vertex which is needed for D$\Gamma$A, at least on a $k$-grid and with a restriction to the most important correlated orbitals.

\begin{figure}
\begin{center}
\includegraphics[clip=true,width=13cm]{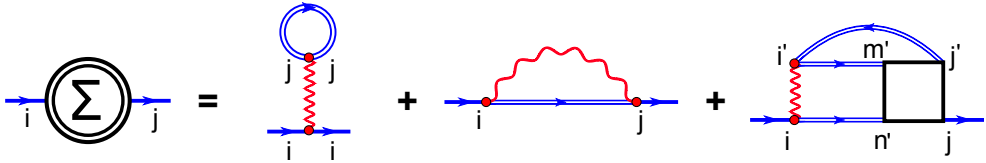}
\end{center}
\caption{{\em Heisenberg equation of motion}  (Schwinger-Dyson eq.) connecting the reducible vertex
  (empty square box) and the self energy \label{Fig:Heisenberg}.}
\end{figure}

The so-calculated reducible vertex $\Gamma$ in turn, is directly related to the self-energy $\Sigma$
through the Heisenberg equation of motion (Schwinger-Dyson eq.) \cite{Toschi06a} (Figure \ref{Fig:Heisenberg}; Hartree and Fock term need to be added):
\begin{equation}
\Sigma_{{\mathbf k}\nu}= - {T^2} \sum_{\stackrel{\nu' \omega}{{\mathbf k'}{\mathbf q}}}  V_{\mathbf q} G_{{\mathbf k+\mathbf q}\, \nu+\omega}G_{{\mathbf k'}\nu'}G_{{\mathbf k'+\mathbf q}\,\nu'+\omega}  \Gamma_{{\mathbf k}\,{\mathbf k'}\, {\mathbf q}}^{\nu\, \nu' \, \omega} .
\label{Eq:Heisenberg}
\end{equation}
Here, $V_{{\mathbf q}}$ denotes the Fourier-transform of $V_{ij}$ which - due to translational symmetry - only depends  on the difference $|i-j|$; $G_{{\mathbf k}\,\nu}$ is the Green function at  momentum $\mathbf k$
and fermionic Matsubara frequency $\nu$; $\omega$ is a bosonic Matsubara frequency; and $T$ denotes the temperature.

\subsection{Inclusion of $GW$ diagrams}
\label{Sec:GW}
Hitherto D$\Gamma$A has been applied only to a simple
Hubbard model with entirely local interaction. As already indicated
in Figure \ref{Fig:DGA} the most natural way to non-local interactions
is to include only the bare Coulomb interaction, and assume
the locality for all  higher
order Feynman diagrams for the
fully irreducible vertex as in Eq.\ (\ref{Eq:Gamma}).
 This way the RPA screening of $GW$
is easily recovered when neglecting the particle-particle channel ($C$) and one of the particle-hole channels ($Z$).
This can be seen from Figure \ref{Fig:parquet} if we consider 
the last particle-hole (interaction) channel, i.e., $\tilde Z$, only:
The first term of the right hand side yields
exactly the RPA relation  between   screened interaction
($\tilde Z$) and  bare one (filled box, fully irreducible vertex) 
if we replace the reducible
vertex $\Gamma$ (empty square) by  $\tilde Z$. The latter replacement 
follows from the first line of Figure \ref{Fig:DGA}, neglecting the
$Z$ and $C$ terms.
Inserting this screened $\tilde Z$ as part of the reducible vertex
into the Heisenberg equation of motion (Figure \ref{Fig:Heisenberg})
yields the screened exchange diagram of $GW$. In other words, if we would
only take the bare Coulomb interaction for the fully irreducible vertex 
{\em and} restrict ourselves to the interaction channel, we recover the
$GW$ approach. 

In realistic calculations with many orbitals it might not be
possible to include all local vertex corrections to the fully irreducible
vertex or to solve the corresponding parquet equations for all orbitals. In this case, it is most reasonable to include these vertex
corrections only for the more strongly interacting $d$- and $f$-electrons,
whereas for the other orbitals only the bare interaction and (possibly) only 
the 
interaction channel is taken into account. That is for these orbitals, only
the $GW$ self energy would be included.

\subsection{Inclusion of DMFT diagrams and non-local correlations beyond}
\label{Sec:DMFT}
The DMFT diagrams, on the other hand, are obtained if we restrict the Green function lines in the parquet equations (Figure \ref{Fig:parquet}) and in the Heisenberg equation of motion (Figure \ref{Fig:Heisenberg}) to their local part.
In this case, the exact set of parquet equations and the exact
Heisenberg equation
of motion connect the local fully irreducible vertex to the
 irreducible vertices in certain channels, to the reducible
vertex and finally the self energy. This way, we have through
exact relations generated all local Feynman diagrams for the 
self energy which is nothing but the DMFT self energy.
These local DMFT diagrams account for the major part of electronic correlations,
which is, among others, responsible for quasiparticle renormalizations including kinks \cite{Nekrasov05a,Byczuk06a,Toschi09a}, Mott-Hubbard transitions \cite{Georges92a}, and  magnetism \cite{Jarrell92a,VollhardtZP,Held98a} with related
spin-polaron physics \cite{Strack92a,Sangiovanni05}.

At lower temperatures and finer energy scales, more subtle effects of electronic correlations emerge, such as spin fluctuations, quantum criticality, and unconventional superconductivity. Such non-local correlations can be described
if we include
the non-local Green functions of {\em ab initio} D$\Gamma$A, i.e., non-local lines in Figures \ref{Fig:parquet} and \ref{Fig:Heisenberg}. Often such physics can be
described in weak coupling perturbation theory. For example,
spin fluctuations are understood as the Bethe-Salpeter ladder in terms of the
bare Coulomb interaction \cite{magnons1,magnons2}, similarly quantum criticality for weakly correlated magnetic systems \cite{Hertz,Moriya,Millis}. Weak localization \cite{Altshuler} and superconductivity \cite{Bardeen57}
on the other hand can be described by the Bethe-Salpeter ladder diagrams
of the particle-particle (Cooperon) channel. All this physics is
included in {\em ab initio} D$\Gamma$A but in terms of the strongly 
renormalized fully irreducible local vertex beyond the bare Coulomb interaction.
Hence these phenomena previously only well described diagrammatically
or weakly correlated electron system can, with D$\Gamma$A,
now be treated for strongly correlated electron systems.

\subsection{{\em Ab initio} D$\Gamma$A algorithm}
\label{Sec:algorithm}
\begin{figure}
\begin{center}
\includegraphics[clip=true,width=7cm]{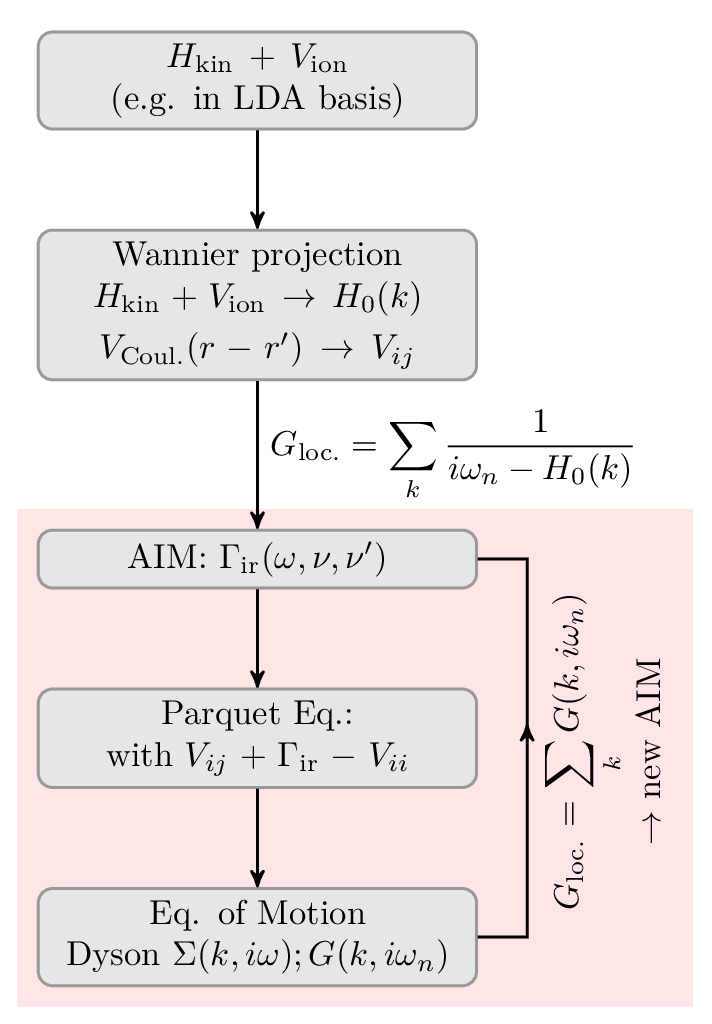}
\end{center}
\caption{{\em Flow diagram of the ab initio D$\Gamma$A} algorithm.
\label{Fig:flow}}
\end{figure}

Let us now, for an overview of the proposed approach, present the flow diagram
of the {\em ab initio} D$\Gamma$A algorithm, see Fig.\ \ref{Fig:flow}.  Starting point is
 an appropriate, material-adjusted basis, e.g., obtained by LDA.
This LDA calculation is done with the full
{\em ab initio} Hamiltonian, 
consisting of kinetic energy  $H_{\rm kin}$,
lattice potential $V_{\rm ion}$, and Coulomb interaction $V_{\rm Coul.}$ within the approximative
exchange correlation potential, see, e.g., \cite{Held07a}. For reducing the 
basis set as well as to identify the more correlated $d$- or $f$-orbitals
on given lattice sites, a Wannier function projection is generally
necessary, e.g., using Wien2Wannier \cite{Wien2Wannier}.
 Within this Wannier basis, kinetic and lattice potential
 are identified as a non-interacting Hamiltonian $H_0$; and
the Coulomb interaction matrix elements $V_{ij}$ are calculated, which
are  treated more sophisticatedly 
in the following D$\Gamma$A calculation. Let us emphasize that the LDA
here only serves to provide an appropriate basis, the LDA approximation itself
is not employed for the actual calculation. 

With this {\em ab initio} starting point, the D$\Gamma$A self consistency cycle is started. From $H_0(\mathbf k)$ the 
local Green function $G_{\rm loc}$ is calculated (in the first iteration
 without self energy). The local interaction $V_{ii}$ and  Green function $G_{\rm loc}$  in turn determine the fully irreducible vertex $\Gamma_{\rm ir}$. 
In practice, $\Gamma_{\rm ir}$ 
is not calculated via summing up Feynman diagrams but by solving (numerically) an Anderson
impurity model (AIM) which has exactly the same local contribution of Feynman diagrams. The numerical effort for calculating this local vertex grows
$\sim n_{\omega}^4$  where  $n_{\omega}$
is the number of Matsubara frequencies. (The vertex has to be calculated for
three Matsubara frequencies and in a Monte-Carlo simulation one needs
to test each corresponding time slice once - to get an independent configuration. There a big prefactor of about ${\cal O}(10^4-10^7)$, depending on the Monte Carlo accuracy needed.)

How to extract the fully irreducible vertex from measured two-particle Green functions has already been discussed in \cite{Toschi06a}. The
fully irreducible vertex for the two dimensional Hubbard model was even calculated for a cluster, see \cite{Jarrellvertex}. Remarkably, the fully irreducible DCA vertex \cite{Jarrellvertex}  is barely $k$-dependent which  strongly supports  our approximation
of a purely local vertex even in two dimensions.

After having calculated $\Gamma_{\rm ir}$, one has to solve the corresponding parquet equations
to get the reducible vertex. The effort of this step grows  $\sim  n_k^4 n_{\omega}^4$
($n_k$:
number of $k$-point patches; $n_{\omega}$:  number of Matsubara frequencies). Hence, for a fine $k$-grid this step becomes the computationally most demanding one.  One might however restrict certain orbitals or larger distances
 to specific channels, i.e., to
the simpler Bethe-Salpeter equation.
This reduces the effort \cite{Toschi06a,Katanin09a} to 
$n_k^2n_{\omega}^4$.
This simpler approach is also  very much in the
spirit of Hedin's original work \cite{Hedin}, emphasizing the interaction-channel, which might be of particular importance for the calculation of actual materials.

From the reducible vertex in turn the self energy is obtained through the Heisenberg equation of motion (\ref{Eq:Heisenberg}). Subsequently, also the  $k$-dependent
Green function and the local Green function are calculated via the usual Dyson equation. The local Green function in turn
allows us to redefine an AIM to recalculate the fully irreducible vertex and to continue with the iteration until convergence.

\section{Summary and outlook}
\label{Sec:outlook}
Twenty-two years after the invention of DMFT
and fourteen years after its merger with LDA we dare to 
look into prospective future developments in the field of
realistic calculations of strongly correlated materials.
We hold that {\em ab initio} D$\Gamma$A, which includes
the bare non-local Coulomb interaction and all local (fully irreducible)
vertex corrections beyond, has a huge potential as a 
purely diagrammatic approach in this respect.
It does not only include $GW$ and DMFT in a natural, unifying way but also 
non-local correlations beyond. 
Hence, it can describe fascinating phenomena of electronic correlations
such as antiferromagnetic spin fluctuations,
their interplay with superconductivity, quantum criticality and weak
localization corrections to the conductivity. Actually, most if not 
all phenomena of strongly correlated electrons we know appear 
to be included, at least in principle, via the corresponding Feynman
diagrams:  {\it ab initio} D$\Gamma$A includes the same kind of Feynman diagrams
used in weak coupling perturbation
theories and ladder summations but with the fully
irreducible vertex instead of only the bare Coulomb interaction.
This way strong electronic correlations are accounted for.
Most of the aforementioned phenomena still need to be explored
by  D$\Gamma$A, as application hitherto concentrated
on spin-fluctuations leading to pseudogaps and the Mermin-Wagner theorem
in two dimensions \cite{Katanin09a} and proper critical exponents in three dimensions \cite{Rohringer11a}.

Despite the advantages, the computational effort nonetheless is still manageable,
possibly with some compromises concerning the number of $k$-points 
for which the full parquet equations instead of the simpler Bethe-Salpeter
equation is solved, and likely with a restriction of the vertex corrections to
the more strongly interacting $d$ - or $f$-orbitals.
Of course, it still needs to be seen how good this description
is in practice for real materials. 
Notwithstanding, we feel that implementing  
{\em ab initio} D$\Gamma$A  is worthwhile the considerable effort to do so.

\subsection*{Acknowledgments}
We thank Philipp Hansmann for  help with the figures
and discussions.
This work was supported  in part by the EU-Indian network MONAMI (GR), by the Austrian science fund (FWF) through
   FOR 1346 (KH)   and 
the Austrian-Russian joint project I-610-N16 (FWF,AT)/10-02-91003-ANF\_a (Russian Basic Research Foundation, AK).

\end{document}